\documentstyle[11pt]{article}
\def\ft#1#2{{\scriptstyle {#1 \over #2}}}
\def\w#1#2{{$W_{#1,#2}$}}
\def\ww3{{$W_3$}}

\def\del{\partial}
\def\a{\alpha}
\def\hk#1{{\hat k_#1}}
\def\dfi{{\del\varphi}}
\def\teff{{T^{\rm eff}}}
\def\cmin{{c_{\rm min}}}
\def\ket#1{{| #1 \rangle}}

\begin{document}
\topmargin 0pt
\oddsidemargin 5mm
\begin{titlepage}
\begin{flushright}
CTP TAMU-8/94\\
LZU-TH-94/02\\
hep-th/9402133\\
February 1994
\end{flushright}
\vspace{1.5truecm}
\begin{center}
{\bf {\Large A note on \w2s strings}}
\vspace{1.5truecm}

{H. Lu,\ \  C.N. Pope\footnote{Supported in part by the U.S.
Department of Energy, under grant DE-FG05-91-ER40633},\ \ X.J. Wang \ and
\ \ S.C. Zhao\footnote{Permanent address: Institute of Theoretical Physics,
Lanzhou University, P.R. China}}
\vspace{1.1truecm}

{\small Center for Theoretical Physics, Texas A\&M University,
                College Station, TX 77843-4242}


\end{center}
\vspace{3.0truecm}
\centerline{\bf{\large ABSTRACT}}
\vspace{1.5truecm}
     BRST operators for two-dimensional theories with spin-2 and spin-$s$
currents, generalising the $W_3$ BRST operator of Thierry-Mieg,
have previously been obtained.  The construction was based on demanding
nilpotence of the BRST operators, making no reference to whether or not an
underlying $W$ algebra exists.  In this paper, we analyse the known cases
($s=3$, 4, 5 and 6), showing that the two $s=4$ BRST operators are
associated with the $W\!B_2$ algebra, and that two of the four $s=6$ BRST
operators are associated with the $W\!G_2$ algebra.  We discuss the
cohomology of all the known higher-spin BRST operators, the Weyl symmetry of
their physical states, and their relation with certain minimal models.
We also obtain the BRST operator for the case $s=7$.

\end{titlepage}
\newpage
\pagestyle{plain}
\section{Introduction}

        In 1987 Thierry-Mieg \cite{mieg} constructed a BRST operator for the
Zamolodchikov $W_3$ algebra \cite{zamo}. The new feature of this BRST
operator, as compared with the one for the standard Virasoro algebra, is that
the $W_3$ algebra is nonlinear.  This nonlinearity is reflected in the fact
that the matter currents are present in the ghost currents. Subsequently it
was shown by Romans \cite{romans} that all scalar realisations of the \ww3
algebra can be expressed in terms of an energy-momentum tensor $T^{\rm eff}$
together with a scalar field $\varphi$ which commutes with $T^{\rm eff}$.
Later it was shown that by performing a nonlinear redefinition involving
$\varphi$ and the spin-2 and spin-3 ghost fields, the BRST operator could be
simplified and written in the form $Q_B=Q_0+Q_1$, where  $Q_0$ has grade
$(1,0)$ and $Q_1$ has grade $(0,1)$, with $(p,q)$ denoting the grading of an
operator with ghost number $p$ for the redefined spin-2 $(b,\,c)$ ghost
system and ghost number $q$ for the redefined spin-3 $(\beta,\,\gamma)$
ghost system \cite{lpsw}. In particular $Q_1$ only involves
$\varphi,\,\beta,\,\gamma$.
This leads to an immediate generalisation \cite{w2s}, in which the BRST
operator has similar form except that the
$(\beta,\,\gamma)$ system is that for a spin-$s$ current rather than a spin-3
current.  In \cite{w2s}, two different BRST operators were found in the case
$s=4$; one in the case $s=5$; and four in the case $s=6$. In this paper we
shall investigate the properties of these various BRST operators in some
detail, including a discussion of their cohomologies and their relation with
certain minimal models. We also extend the previous results to the case
$s=7$, and find that there is just one BRST operator for this case.

 The BRST operator for the spin-2 plus spin-$s$ string takes the form
\cite{w2s}:
\begin{eqnarray}
Q_B&= &Q_0 + Q_1,\label{eq1}\\
Q_0&=&\oint dz\, c \Big(T^{\rm eff} +T_{\varphi} + T_{\gamma,\beta} + \ft12
T_{c,b} \Big), \\
Q_1&=&\oint dz\, \gamma \, F(\varphi,\beta,\gamma),
\label{eq3}
\end{eqnarray}
where the energy-momentum tensors are given by

\begin{eqnarray}
T_\varphi & =&-\ft12 (\del\varphi)^2 -\alpha\, \del^2\varphi,\\
T_{\gamma,\beta} &=& -s\, \beta\,\del\gamma -(s-1)\, \del\beta\, \gamma,
\\
T_{c,b} & =&  -2\, b\, \del c - \del b\, c,
\\
T^{\rm eff} &= &-\ft12  \eta_{\mu\nu}\,\del X^\mu\, \del X^\nu\ -
i a_\mu\, \del^2 X^\mu.
\label{eq7}
\end{eqnarray}

The operator $F(\varphi,\beta,\gamma)$ has spin $s$ and ghost number zero.
Because of the grading discussed above, it follows that one will
have the nilpotency conditions $Q_0^2=Q_1^2=\{Q_0,Q_1\}=0$.  The first of
these conditions is satisfied provided that the total central charge
vanishes, {\it i.e.}\
\begin{equation}
0=-26 -2(6s^2-6s+1) + 1+12\alpha^2 + c^{\rm eff}\ ,
\end{equation}
where $c^{\rm eff}$ is the central charge for $\teff$.
The remaining two nilpotency conditions determine the precise form of the
operator $F(\varphi,\beta,\gamma)$ appearing in (\ref{eq7}).  Solutions for
$s=$4, 5 and 6 were found in \cite{w2s}. Together with the $s=7$ case which
we find in this paper, the list of known $W_{2,s}$ BRST operators is
\vskip20pt
\begin{table}
\begin{tabular}{l|p{1.25in} p{1.25in} p{1.25in}} \hline\hline
\phantom{theory} &$\alpha^2$ &$a^2$ &$c_{\rm min}$ \\
\hline
\w23 &$\ft{49}{8}$ &$\ft{49}{24}$ &$\phantom{-}\ft12$\\
\hline
\w24 &$\ft{243}{20}$ &$\ft{121}{60}$ &$\phantom{-}\ft45$\\  \cline{2-4}
\phantom{w24} &$\ft{361}{30}$ &$\ft{32}{15}$ &$-\ft35$\\
\hline
\w25 &$\ft{121}{6}$ &2 &$\phantom{-}1$\\
\hline
\w26 &$\ft{845}{28}$ &$\ft{167}{84}$ &$\phantom{-}\ft87$\\  \cline{2-4}
\phantom{w26} &$\ft{1681}{56}$ &$\ft{361}{168}$ &$-\ft{11}{14}$\\ \cline{2-4}
\phantom{\w26} &$\ft{5041}{168}$ &$\ft{121}{56}$ &$-\ft{13}{14}$\\ \cline{2-4}
\phantom{\w26} &$\ft{361}{12}$ &$\ft{25}{12}$ &$\phantom{-}0$ \\
\hline
\w27 &$\ft{675}{16}$ &$\ft{95}{48}$ &$\phantom{-}\ft54$\\ \hline\hline
\end{tabular}
\label{table1}
\caption{The \w23--\w27 BRST operators}
\end{table}
\baselineskip=15pt

\vskip20pt
In the above table, $c_{\rm min}$ is the central charge of
$T_{\varphi}+T_{\gamma,\beta}$, and is related to $c^{\rm eff}$ by $c_{\rm
min}=26-c^{\rm eff}$.  The column labelled by $a^2$ gives the value of the
background charge for $T^{\rm eff}$ in the case where it is realised in
terms of a single scalar field $X$:
\begin{equation}
T^{\rm eff}=-\ft12 (\del X)^2 -a\, \del^2 X \, .
\end{equation}

For each $s$, the first BRST operator listed in the above table corresponds to
a
general sequence of BRST operators with $\alpha^2=(s-1)(2s+1)^2/(4(s+1))$ and
$c_{\rm min}=2(s-2)/(s+1)$. In these cases, as discussed in \cite{wmod}, the
$(\varphi,\,\beta,\,\gamma)$ system provides a realisation of the lowest
non-trivial unitary minimal model for the $W_{s-1}$ algebra. In the case of
a multi-scalar realisation for $T_{\rm eff}$, this sequence of BRST operators
is associated with a unitary string theory. In this paper, the principal
focus will be on the remaining exceptional BRST operators. Although in a
multi-scalar realisation these would in general correspond to non-unitary
string theories, they nevertheless provide explicit realisations for certain
minimal models.

In section 2 we study the \w24 BRST operators, paying particular attention
to the second case with $\a^2=\ft{361}{30}$. We shall study its relation to
the (3,5) Virasoro minimal model, the Weyl symmetry of its physical states,
and the complete cohomology of its BRST operator, using the method developed
in \cite{cohom}. In section 3 we study the $s=6$ BRST operators, especially
the last three \w26 BRST operators in table \ref{table1}. We shall study
their connection to minimal models, the Weyl symmetry of the physical
states, and comment on their physical spectra.  In section 4 we bosonise the
spin-$s$ ghosts, and discuss the new features of the BRST operators. The
paper ends with a summary.

\section{The \w24 BRST operators}

     In general, the $W_{2,s}$ BRST operators are not associated with closed
algebras of spin-2 and spin-$s$ currents at the quantum level.  However,
there is such an association in certain special cases, such as $W_{2,3}$
({\it i.e.}\ the usual $W_3$ algebra).  There are two further examples where
there are underlying $W$ algebras, namely for $s=4$, which we shall discuss
in this section, and $s=6$, which will be discussed in section 3.
These BRST operators were all obtained by imposing the requirement of
nilpotence on the ansatz in equations (\ref{eq1})--(\ref{eq7}), without
making any reference to any possible underlying $W$ algebra.  It is in fact
not easy to extract the currents of such an algebra from this form of the
BRST operator, and so in order to uncover the relation with the $W$ algebra
we find it simpler to adopt a more indirect approach.  This was done already
for the \w24 BRST operator with $\alpha^2=\ft{243}{20}$ in \cite{zhao},
where we showed that the physical states in the two-scalar case form
representations under the $B_2$ Weyl group.  This indicates that the BRST
operator is one for the $W\!B_2$ algebra, which is generated by an
energy-momentum tensor and a spin-4 primary current.

     We shall now examine the second \w24 BRST operator, with
$\alpha^2=\ft{361}{30}$.  This is given by (\ref{eq1})--(\ref{eq7}), with
$s=4$ and
\begin{eqnarray}
F(\beta,\gamma,\varphi)&=&(\del\varphi)^4 + 4\alpha\, \del^2\varphi\,
(\del\varphi)^2 + \ft{253}{30} (\del^2\varphi)^2 + \ft{39}{5}
\del^3\varphi\, \del \varphi +\ft{41}{570} \alpha\, \del^4\varphi\nonumber\\
& \phantom{=} & + 8 (\del\varphi)^2\, \beta\, \del\gamma
-\ft{26}{19}\alpha\, \del^2\varphi\, \beta\, \del\gamma -\ft{66}{19}
\alpha\, \del\varphi\, \beta\, \del^2 \gamma  \nonumber\\
& \phantom{=} &  - \ft{26}{15} \beta\, \del^3\gamma + \ft{29}5
\del^2\beta\, \del\gamma .
\label{w24f}
\end{eqnarray}
Physical states are defined in the usual way by the requirement that they be
BRST closed, but not exact.  Using the method developed in \cite{cohom}, we
shall now discuss this cohomology problem for the two-scalar case.

     From the pattern of the explicit low-level physical states that we have
obtained, we have observed that, just as in the case of the two-scalar $W_3$
string, the momenta $(p_1,\, p_2)$ of all physical states are quantised in
rational multiples of the background charges $(\alpha,\, a)$ for the scalars
$(\varphi,\, X)$.  Specifically, in this case, we find
\begin{equation}
p_1={k_1\over19}\, \alpha,\qquad\qquad p_2={k_2\over 8}\, a\, ,
\end{equation}
where $k_1$ and $k_2$ are integers.  All physical states satisfy a
mass-shell condition,
\begin{equation}
5(12\ell+1)=(k_1+19)^2 +(k_2+8)^2\, ,
\label{ms24}
\end{equation}
where $\ell$ is the level number.

     The method used in \cite{cohom} to solve the cohomology consists of
finding two particular physical operators, which we shall call $x$ and $y$,
which have the property that they have inverses, $x^{-1}$ and $y^{-1}$, in
the sense that the normal-ordered products $(x^{-1}\, x)$ and $(y^{-1}\, y)$
are non-vanishing constants.  These operators have the further property that
they have well-defined normal-ordered products with all physical operators,
and so they map physical operators into other BRST non-trivial physical
operators. They may therefore be used to generate the entire cohomology from
a basic set of physical operators whose momenta lie in a fundamental unit
cell.

      The simplest candidates for the $x$ and $y$ operators in the present
case have momenta ($k_1,\,k_2$) given by
\begin{eqnarray}
x& : & (15,\,15)\qquad\qquad x^{-1} : \ (-15,\,-15) \nonumber\\
y&:  & (15,\,-15) \qquad\quad y^{-1}: \ (-15,\,15)
\end{eqnarray}
It is easy to see from the mass-shell condition (\ref{ms24}) that operators
with these momenta have well-defined normal-ordered products with those for all
integer solutions of (\ref{ms24}), and hence in particular with all physical
operators. One can also see from (\ref{ms24}) that $x$ and $x^{-1}$ have
level numbers $\ell=28$ and $\ell=1$; whilst $y$ and $y^{-1}$ have level
numbers $\ell=20$ and $\ell=9$. We have been able to construct the operators
$x^{-1}$ and $y$ explicitly. We find that the $x^{-1}$ operator has ghost
number $G=3$, which implies that the $x$ operator has $G=-3$. The $y$
operator has ghost number $G=-1$, which implies that $y^{-1}$ has $G=1$.
Although we have been unable to construct the $x$ and $y^{-1}$ operators
explicitly, owing to their complexity, we have accumulated considerable
evidence that supports their existence and invertibility. Assuming
for now that this is the case, one can easily see that the fundamental cell
may be taken as a $30\times 15$ rectangle in the ($k_1,\,k_2$)-plane. There
are 40 integer solutions to the mass-shell condition (\ref{ms24}) in any
such rectangle, of which 36 turn out to correspond to physical states.  The
remaining four, which can be characterised by $k_1=1,\,6\, {\rm mod}\ 15$,
do not correspond to physical states. For our calculations we chose the
fundamental cell to be the rectangle with $-34\le k_1\le -5$, $-15\le k_2\le
-1$.  All the physical operators in this cell have level numbers $\ell\le 4$
and it is straightforward to find them explicitly. The complete cohomology
can be obtained  by normal-ordering $x^m\,y^n$ with all the physical
operators in this fundamental cell, for arbitrary integers $m$ and $n$.

   Although, as we remarked above, we were unable to construct explicitly
the $x$ and $y^{-1}$ operators, we have found numerous examples of pairs of
physical states whose momenta and ghost numbers are consistent with their
existence as cohomology generating operators.

      We now turn to the consideration of the $B_2$ Weyl group and its
action on physical states. For the \w24 BRST operator with
$\a^2=\ft{243}{20}$, this was discussed extensively in \cite{zhao}. For the
present case, with $\a^2=\ft{361}{30}$, we first define the shifted momenta
$\hat k_1=k_1+19$, and $\hat k_2=k_2+8$. The $B_2$ Weyl group is generated
by
\begin{eqnarray}
S_1\ &:&\qquad (\hat k_1,\, \hat k_2)\longrightarrow (\hat k_1,\, -\hat k_2)
\ ,\nonumber\\
S_2\ &:&\qquad (\hat k_1,\, \hat k_2)\longrightarrow (\hat k_2,\, \hat k_1)
\ .\label{weyl24}
\end{eqnarray}
The eight elements of the Weyl group are given by 1, $S_1$, $S_2$, $S_1S_2$,
$S_2S_1$, $S_1S_2S_1$, $S_2S_1S_2$, $S_1S_2S_1S_2$. It is clear that the Weyl
group leaves the mass-shell condition (\ref{ms24}) invariant. Furthermore it
maps the momentum of any physical operator into the momentum of another
physical operator \cite{cohom}.  For  example, the eight tachyons
$c\del^2\gamma \del\gamma\,\gamma e^{p_1\varphi+p_2 X}$ have momenta
$(\hk1,\,\hk2)=(\pm1,\,\pm2), \ (\pm2,\,\pm1)$, where the $\pm$ signs are
independent.  These manifestly form a multiplet under the Weyl group.

     The existence of this Weyl group symmetry indicates that the BRST
operator is based on the $W\!B_2$ algebra. The explicit realisation of the
$W\!B_2$ algebra is given in \cite{kausch,zhao}. Owing to the fact that
$B_2$ is a non-simply-laced algebra, there are two inequivalent choices of
the background charge for $\varphi$ that give the same critical value of the
central charge.  As was shown in \cite{zhao}, these two values are
$\a^2=\ft{243}{20}$ and $\a^2=\ft{361}{30}$; {\it i.e.}\ precisely the two
values that were found for the \w24 BRST operators.

         In previous studies of $W$ string theories, it has been found that
the physical states in the case of a multi-scalar realisation can be
associated with the primary fields of certain minimal models.  To be more
specific, the physical operators take the form
\begin{equation}
c\,U(\varphi,\beta,\gamma) \,R(X^\mu)\ ,
\label{form}
\end{equation}
where $U(\varphi,\beta,\gamma)$ is a primary field of the
minimal model, and $R(X^\mu)$ is a highest weight state under $T^{\rm eff}$.
The cohomology of a multi-scalar $W$ string can be associated with a subset
of the cohomology of the two-scalar $W$ string in a way that has been
extensively discussed in \cite{cohom,zhao}.  In the case of the \w24 string
with $\a^2=\ft{361}{30}$, we find that the weights of primary operators
$U(\varphi,\beta,\gamma)$  are $h=(-\ft1{20},\,0,\,\ft15,\,\ft34)$. Noting
that the central charge of the $(\varphi,\beta,\gamma)$ system is $c_{\rm
min}=-\ft35$, we recognise that the minimal model in this case is
the $(3,\,5)$ Virasoro minimal model.

\section{The \w26 BRST operators}

     There is just one further $W_{2,s}$ algebra that exists for arbitrary
central charge (and, in particular, for the value that is needed for
criticality), namely $W\!G_2$, which is generated by an energy-momentum
tensor and a spin-6 primary current \cite{stan}.  Since $G_2$ is non-simply
laced, there will again be two inequivalent realisations, with different
background charges, that give rise to nilpotent BRST operators.  The form of
the energy-momentum tensor is
\begin{equation}
T=-\ft12(\del\vec\varphi)^2 - \big(t\, \vec\rho +{1\over t}\,
\vec\rho^{\,\vee}
\big) \cdot \del^2\vec \varphi\, ,
\end{equation}
where $\vec\rho=(\ft32,\, \ft{\sqrt3}{6})$ is the $G_2$ Weyl vector and
$\vec\rho^{\,\vee}=(5,\, \sqrt3)$ is the co-Weyl vector.  Solving for the
critical central charge $c=388$ gives two solutions for $t^2$, namely
$t^2=\ft72$ and $t^2=\ft{24}7$.  These correspond to a background charge
$\a$ given by $\a^2=\ft{1681}{56}$ and $\a^2=\ft{5041}{168}$.  Thus we see
that the second and third \w26 BRST operators listed in table \ref{table1}
correspond to BRST operators of the $W\!G_2$ algebra.  We shall demonstrate
below that indeed the physical states of these two BRST operators display a
symmetry under the action of the Weyl group of $G_2$.  The other two \w26
BRST operators, with $\a^2=\ft{845}{28}$ and $\a^2=\ft{361}{12}$, are not
associated with any underlying $W$ algebra, and in fact as we shall show,
there is no Weyl-group symmetry in these cases.

     First, we shall consider the two BRST operators associated with the
$W\!G_2$ algebra.  The low-level physical spectra of the two-scalar
realisation indicate that the momenta of all physical states are quantised
in rational multiples of the background charges $\a$ and $a$.  For the case
$\a^2=\ft{5041}{168}$, we find that the momenta have the form
\begin{equation}
p_1=\Big({\hk1\over 71}-1\Big)\, \a\ , \qquad\qquad p_2=\Big({\hk2\over 11}-1
\Big)\, a\ ,
\end{equation}
where $\hk1$ and $\hk2$ are integers.  (We find it convenient to work directly
with the shifted momentum variables here.)  The mass-shell condition in this
case is
\begin{equation}
28(12\ell+1)=\hk1^2+3\hk2^2\, .
\end{equation}
It is easy to see that this equation is invariant under the twelve-element
$G_2$ Weyl group generated by
\begin{eqnarray}
S_1&:& (\hk1,\, \hk2)\longrightarrow (\hk1,\, -\hk2)\, ,\nonumber\\
S_2&:& (\hk1,\, \hk2)\longrightarrow \big(\ft12(-\hk1 +3\hk2),\,
\ft12(\hk1+\hk2)\big)\, .
\end{eqnarray}
We find that the tachyons, which are given by $c\, \del^4\gamma\,
\del^3\gamma\, \del^2\gamma\, \del\gamma\, \gamma\, e^{p_1\varphi+p_2 X}$,
have momenta $(\hk1,\, \hk2)=$ $(\pm1,\, \pm 3)$, $(\pm 4,\, \pm 2)$ and
$(\pm 5,\, \pm1)$, where the $\pm$ signs are independent. These twelve
momenta of tachyons do indeed form a multiplet under the Weyl group.

     For the \w26 BRST operator with $\a^2=\ft{1681}{56}$, we find that the
momenta of all physical states take the form
\begin{equation}
p_1=\Big({\hk1\over 41}-1\Big)\, \a\ , \qquad\qquad p_2=\Big({\hk2\over 19}
-1\Big)\, a\ ,
\end{equation}
where $\hk1$ and $\hk2$ are again integers.  The mass-shell condition is
\begin{equation}
28(12\ell+1)=3\hk1^2+\hk2^2\, .
\end{equation}
This is invariant under the twelve-element $G_2$ Weyl group generated by
\begin{eqnarray}
S_1&:& (\hk1,\, \hk2)\longrightarrow (\hk1,\, -\hk2)\, ,\nonumber\\
S_2&:& (\hk1,\, \hk2)\longrightarrow \big(\ft12(\hk1+\hk2),\,\ft12(3\hk1 -\hk2)
\big)\, .
\end{eqnarray}
The tachyons here have momenta $(\hk1,\, \hk2)=$ $(\pm 1,\, \pm 5)$, $(\pm
2,\, \pm 4)$ and $(\pm 3,\, \pm 1)$, where the $\pm$ signs are independent.
The momenta of these twelve tachyons again form a multiplet under the Weyl
group.

     We expect that these results should generalise to higher-level physical
states.  In order to investigate this completely, one would like to find the
cohomology-generating operators $x$, $x^{-1}$, $y$ and $y^{-1}$ in each
case.  We have not attempted to do this here, because of the complexity of
the BRST operators.  However, it is very plausible that they exist.

     Like the $W$ strings previously studied, we would expect that the
physical states of the multi-scalar $W\!G_2$ strings have the form given
in (\ref{form}).   The physical states of this form are relatively easy to
construct, and we have studied all the low-level examples for both the BRST
operators of $W\!G_2$.   For the case with $\a^2=\ft{5041}{168}$, we find
that the weights of the primary operators $U(\varphi, \beta, \gamma)$ are
given by $h=(-\ft1{14},\, -\ft5{112}, \, 0,\, \ft17,\, \ft{27}{112},\,
\ft9{14},\, \ft{13}{16},\, \ft{10}7,\, \ft52)$.  The central charge of the
$(\varphi, \beta, \gamma)$ system is $c_{\rm min} =-\ft{13}{14}$.  These
weights and central charge are precisely those of the $(4,7)$ Virasoro
minimal model.   It is interesting to note that this value of central charge
is exactly the one found by Zamolodchikov to be necessary for the
associativity of the $W_{2,\ft52}$ algebra, which is generated by the
energy-momentum tensor and a spin-$\ft52$ current \cite{zamo}.

     For the case with $\a^2=\ft{1681}{56}$, the central charge of the
 $(\varphi, \beta, \gamma)$ system is $c_{\rm min} =-\ft{11}{14}$.  We can
expect in this case that the operators $U(\varphi,\beta,\gamma)$ should
correspond to the primary fields of the $(7,12)$ Virasoro minimal model.  In
fact this central charge is also equal to that for one of the $W\!B_2$ minimal
models.  At level $\ell=19$ and $G=1$, there exists a physical state with
momentum $p_1=0$, whose operator $U(\varphi,\beta,\gamma)$ has weight $h=4$,
given by
\begin{eqnarray}
U &=& (\del\varphi)^4 +4\a\, \del^2\varphi (\dfi)^2 +\ft{21557}{532}
(\del^2\varphi)^2 + \ft{425}{38} \dfi\del^3\varphi +\ft{727}{9348} \a\,
\del^4\varphi  \nonumber\\
&\phantom{=} &+ 24 (\dfi)^2\,\beta\,\del\gamma +\ft{2154}{133} (\dfi)^2
\del\beta\,\gamma +\ft{3036}{133} \dfi\del^2\varphi\,\beta\gamma
+\ft{30360}{5453} \a\,\dfi\,\del\beta\,\del\gamma
 \nonumber \\
&\phantom{=} &+\ft{18975}{5453} \a\,\dfi\,\del^2\beta\gamma
+\ft{79584}{5453}\a\, \del^2\varphi\beta\,\del\gamma
+\ft{11679}{779} \a \,\del^2\varphi\,\del\gamma\,\gamma
+\ft{1518}{779}\a\, \del^3\varphi\,\beta\gamma   \nonumber\\
&\phantom{=} & -\ft{1325}{133} \beta\,\del^3\gamma
+ \ft{2650}{133} \del\beta\,\del^2\gamma
+ \ft{21729}{931} \del^2\beta\,\del\gamma
+ \ft{11205}{1064} \del^3\beta\,\gamma
+ \ft{7632}{133} \del\beta\,\beta\,\del\gamma\,\gamma
\end{eqnarray}
The energy-momentum tensor $T^{\rm min}$ of the $(\varphi,\beta,\gamma)$ system
is given by
\begin{equation}
T^{\rm min} = -\ft12 (\del \varphi)^2 - \a \del^2\varphi -6\beta\,\del\gamma
-5 \del\beta\,\gamma \ .
\end{equation}
We have explicitly verified that currents $T^{\rm min}$ and $U$ provide
a realisation of the $W\!B_2$ algebra at central charge $\cmin=-\ft{11}{14}$
(up to appearance of the BRST trivial terms in the OPE of $U(z)\, U(w)$).
This indicates that the $U(\varphi,\,\beta,\,\gamma)$ operators, which give
a representation of the $(7,12)$ Virasoro minimal model, can be viewed as the
primaries and descendants of the $W\!B_2$ minimal model with
$\cmin=-\ft{11}{14}$. The $(7,12)$ Virasoro minimal model has 33 primaries,
21 of which are primaries of the $W\!B_2$ minimal model, with weights
($-\ft1{14},\,-\ft1{16},-\ft1{21}$,\,$-\ft3{112},\,0,\,\ft{11}{336},$\,
$\ft{13}{112},\,\ft16,\,\ft{25}{112}$,\,$\ft27,\,\ft37,\,\ft{25}{42},$\,
$\ft{11}{16},\,\ft{11}{14},\,
\ft{299}{336},$\,$\ft{153}{112},\,\ft32,$\,$\ft{44}{21},\,
\ft{125}{48},\,\ft{187}{42},\,\ft{23}3)$.
The remaining 12  are $U$ descendants with weights $(\ft{125}{112},\,\ft{17}7,
\,\ft{333}{112},\,\ft{377}{112},$\,$4,\,\ft{25}{14},\,\ft{69}{14},$\,
$\ft{91}{16},\,\ft{697}{112},$\,$\ft{58}7,\,\ft{159}{16},\,\ft{25}2 )$.
This is an explicit example of the phenomenon where the set of highest weight
fields of a $W$ minimal model gives rise to a larger set of highest weight
fields with respect to the Virasoro subalgebra. In this example, although
not in general, the set of Virasoro primaries is finite.

    Finally, we turn to the last of the \w26 BRST operators, with
$\a^2=\ft{361}{12}$. In this case the $(\varphi,\,\beta,\,\gamma)$ system has
central charge $\cmin=0$. This BRST operator does not seem to correspond to
any known algebra. Nevertheless  the low-level examples indicate that the
momenta of physical states are quantised in rational multiple of $\a$ and
$a$. We find that $p_1$ and $p_2$ are given by
\begin{equation}
p_1=\Big({\hk1\over 95}-1\Big)\, \a\ ,\qquad\qquad p_2=\Big({\hk2\over25} -1
\Big)\, a\ ,
\end{equation}
where $\hk1$ and $\hk2$ are integers.  The mass-shell condition in this case
is
\begin{equation}
50(12\ell+1)=\hk1^2 + \hk2^2\ .
\label{mm}
\end{equation}
The twelve tachyons have momenta given by $(\hk1,\, \hk2)=$ $(\pm 5,\, \pm
5)$, $(\pm 1,\, \pm 7)$ and $(\pm 7,\, \pm 1)$.  There is no discrete group
of order 12 that maps these momenta into each other and preserves the
mass-shell condition (\ref{mm}).  Thus we conclude that there is no Weyl
group in this case.  (Note that (\ref{mm}) is similar to (\ref{ms24}), and
is therefore invariant under the $B_2$ Weyl group generated by (\ref{weyl24}).
However, this does not act transitively on the 12 tachyons; those
with momenta $(\pm 5,\, \pm 5)$ are mapped into one another, and the
remaining 8 tachyons form a multiplet amongst themselves.)  In the
multi-scalar case, the weights of the primary operators $U(\varphi,\beta,
\gamma)$ have values including $h=(-\ft1{25},\, 0,\, \ft1{25},\,
\ft4{25},\, \ft{11}{25},\, \ft{14}{25},\, \ft{21}{25})$ \cite{w2s}.  It is
unclear whether these weights can be associated with any ``minimal model''
of the $c_{\rm min}=0$ system realised by $(\varphi,\, \beta,\, \gamma)$.

\section{Bosonisation of the $\beta$, $\gamma$ ghosts}

    It was recently observed in \cite{fw,west} that by bosonising the $\beta$
and $\gamma$ ghosts, so that they are written in terms of a free scalar
field $\rho$,  one can, after performing a non-local transformation
involving $\rho$ and $\varphi$, cast the $W_3$ string into the form of a
parafermionic theory.

     For the general sequence of $W_{2,s}$ BRST operators with
$\a^2=(s-1)(2s+1)^2/(4(s+1))$, one can perform the following bosonisation of
$\beta$ and $\gamma$:
\begin{equation}
\beta=e^{-i\rho}\ , \qquad \qquad \gamma=e^{i\rho}\ .
\end{equation}
After the field redefinition
\begin{eqnarray}
\varphi &=& -i\sqrt{s^2-1}\, \phi_1 + s\, \phi_2 \ ,\nonumber\\
\rho &=& -s\, \phi_1 -i \sqrt{s^2-1}\, \phi_2 \ ,
\label{fw}
\end{eqnarray}
the $Q_1$ BRST operator can be written as
\begin{equation}
Q_1 = \oint dz\, :\big(\del^s e^{i \phi_1}\big) e^{-i (s+1)\phi_1}:
e^{\sqrt{s^2-1} \phi_2}\ .
\label{newQ}
\end{equation}
The case $s=3$ was proved in \cite{west}; we have explicitly verified the
result for the cases of $s=4$, 5 and 6.  In addition, we have constructed
the $W_{2,7}$ BRST operator according to the general ansatz
(\ref{eq1})--(\ref{eq7}) and
verified that in this case also, $Q_1$ can be written in this form.
Note that in (\ref{newQ}) we have indicated the normal ordering for the
$\phi_1$ terms explicitly, to emphasise that there are no contractions
between the fields of the two $\phi_1$--dependent exponentials.

     In \cite{west}, the following non-local transformation from $\phi_1$ to
$\hat \phi_1$ was then performed:
\begin{equation}
e^{i\hat\phi_1}=-\del e^{-i\phi_1}\ ,\qquad\qquad e^{i\phi_1}= \del
e^{-i\hat \phi_1}\ .
\label{dis}
\end{equation}
One can easily verify that under this, (\ref{newQ}) becomes
\begin{equation}
Q_1=\oint dz\, e^{i s\hat\phi_1}\, e^{\sqrt{s^2-1}\phi_2}\ .
\label{pf}
\end{equation}
(In this equation, and in (\ref{newQ}), we have dropped an unimportant
overall constant factor.)

      Although the transformation (\ref{dis}) is canonical, {\it i.e.}\
$\hat\phi_1$ satisfies the same OPE as $\phi_1$, it has a significant effect
on the cohomology of the BRST operator $Q_B$.  In fact, all the original
physical states become BRST trivial.  Actually one can prove an even
stronger result, namely that after the transformation (\ref{dis}) the
cohomology of $Q_B$ becomes trivial.   To see this, it is convenient to
perform a further canonical transformation, from the fields $(\hat\phi_1,\,
\phi_2)$ to $(\tilde\phi_1,\, \tilde\phi_2)$, defined by
\begin{eqnarray}
\tilde\phi_1 &=& s\, \hat\phi_1 - i\sqrt{s^2-1}\, \phi_2\ ,\nonumber\\
\tilde\phi_2 &=& i\sqrt{s^2-1}\, \hat\phi_1 + s\, \phi_2\ .
\label{nt}
\end{eqnarray}
The $Q_1$ BRST operator then becomes simply $Q_1=\oint dz\,e^{i\tilde\phi_1}$.
If we perform the fermionisation $\tilde\beta=e^{-i \tilde\phi_1}$,
$\tilde\gamma= e^{i \tilde\phi_1}$, we may write $Q_B$ as
\begin{eqnarray}
Q_B &=& Q_0+Q_1 \ ,\nonumber\\
Q_0 &=& \oint dz\, c\big(T_m +\del\tilde\beta\, \tilde\gamma +\ft12 T_{c,b}
\big)\ ,\qquad Q_1=\oint dz\, \tilde\gamma \ ,
\label{tr}
\end{eqnarray}
where $T_m=\teff + T_{\tilde\phi_2}$ is a matter energy-momentum tensor with
central charge 28.  Note that $\tilde\gamma$ has spin 1, and $\tilde \beta$
has spin 0.

     To see that $Q_B$ defined in (\ref{tr}) has no non-trivial cohomology,
we write an arbitrary physical state $\ket{\chi}$ in the form $\ket{\chi}=
\tilde\beta \ket{\chi_1} + \ket{\chi_2}$, where $\ket{\chi_1}$ and
$\ket{\chi_2}$ contain no undifferentiated $\tilde\beta$ field.  It is easy to
see that the physical-state condition $Q_B\ket{\chi}=0$ implies that
$\ket{\chi_1}= -Q_0 \ket{\chi_2}$, and hence that $Q_B\, \tilde
\beta\ket{\chi_2}= Q_0 \tilde
\beta\ket{\chi_2} + Q_1 \tilde
\beta\ket{\chi_2}= -\tilde\beta \, Q_0 \ket{\chi_2} +\ket{\chi_2}
=\ket{\chi}$. Thus any state that satisfies the physical-state condition
$Q_B\ket{\chi}=0$ is BRST trivial.  Since the transformation (\ref{nt}), and
the subsequent fermionisation that we performed above, will preserve the
cohomology of $Q_B$, it follows that the original cohomology of $Q_B$ had
already collapsed at the stage of the parafermionic formulation in
(\ref{pf}), owing to the non-locality of the transformation (\ref{dis}).

     It is remarkable that despite the non-locality of the transformation
(\ref{dis}), the original physical operators can apparently all be re-expressed
in terms of the $(\hat\phi_1,\, \phi_2)$ fields.  We have checked this in
several non-trivial examples for the $W_3$ string, for various levels up to
$\ell=14$.  The reason why these operators, which are BRST trivial in the
$(\hat\phi_1,\, \phi_2)$ parametrisation, are nevertheless BRST non-trivial
prior to the transformation (\ref{dis}) is that they are written as $Q_B$
acting on operators built from $(\hat\phi_1,\, \phi_2)$ that cannot
themselves be re-expressed in terms of the original fields, owing to the
non-locality of (\ref{dis}).

     One of the striking features of the transformation (\ref{fw}) is that
for the general sequence of BRST operators that we have just been
discussing, one can, by adding total derivatives, express $Q_1$ in a form
where $\phi_2$ appears only in the exponential, with monomials involving
derivatives only of $\phi_1$ in the prefactor.  Furthermore, all these terms
can then be expressed in the simple form given in (\ref{newQ}).  It is of
interest to see how much of this can be carried through for the remaining
exceptional BRST operators for \w24 and \w26.  We find that the first step,
namely eliminating all the $\phi_2$ dependence from the prefactor, can be
achieved for all the remaining BRST operators.  For example, in the case of
the \w24 BRST operator with $\alpha^2=\ft{361}{30}$, we can perform the
following transformation:
\begin{eqnarray}
\varphi &=& -i \sqrt{120}\, \phi_1 + 11\, \phi_2\ ,\nonumber\\
\rho &=& -11\, \phi_1 -i \sqrt{120}\, \phi_2 \ ,
\end{eqnarray}
leading to
\begin{equation}
Q_1=\oint dz\, \Big( 6\, (\del\phi_1)^4 - i\sqrt{96}\, (\del\phi_1)^2\,
\del^2\phi_1 -13\, (\del^2\phi_1)^2 + \del\phi_1\, \del^3\phi_1 -{i\over
\sqrt{6}}\, \del^4\phi_1\Big) e^{-i\sqrt{6}\phi_1 +\sqrt{5}\phi_2}\ .
\label{w24q1}
\end{equation}
It is not possible to write this in the simple form (\ref{newQ}).  It is
unclear whether there is nevertheless some more general kind of non-local
transformation analogous to (\ref{dis}) that could transform $Q_1$ into a
parafermionic form such as (\ref{pf}).  Our findings for the exceptional
\w26 BRST operators are similar, in that we obtain analogous expressions to
(\ref{w24q1}), which again cannot be expressed in a form such as
(\ref{newQ}).

\section{Summary}

      In this paper, we have studied all the $W_{2,s}$ BRST operators up to
$s=7$, paying particular attention to the execeptional $W_{2,s}$ BRST
operators for $s=4$ and $s=6$ that have not been previously investigated in
any detail.  These BRST operators would describe non-unitary string theories
in multi-scalar realisations.  However, the cohomologies of these BRST
operators in multi-scalar realisation give rise to explicit realisations of
certain minimal models.  In two-scalar realisation, the cohomologies of
these BRST operators are much richer.  Since the momenta of all the
physical states seem to be quantised in rational multiples of the background
charges, we would expect that there exist cohomology-generating operators in
all the exceptional cases, and thus one would obtain the corresponding
complete cohomologies, using the method developed in \cite{cohom}.   In this
paper, we carried out the task for the exceptional $W_{2,4}$ BRST operator.

     We also identified the $W$ symmetries that underlie certain of the
$W_{2,s}$ BRST operators.   The two $W_{2,4}$ BRST operators are associated
with $W\!B_2$ algebra, and two of the $W_{2,6}$ BRST operators are
associated with $W\!G_2$ algebra. None of the other $W_{2,s}$ BRST operators
with $s\ge 5$ are associated with any underlying quantum $W$ algebras.
With the
possible exception of $W_{2,6}$ with $\a^2=\ft{361}{12}$, all the $W_{2,s}$
BRST operators give rise to explicit realisations of certain minimal models.
In particular, $W_{2,6}$ BRST operator with $\a^2=\ft{1681}{56}$ is
associated with $W\!B_2$ minimal model with $c_{\rm min}=-\ft{11}{14}$.  The
$W_{2,6}$ BRST operator with $\a^2=\ft{361}{12}$ has the central charge
$c_{\rm min}=0$.  It is unclear
whether this BRST operator is associated with any ``minimal'' model.

     We also explicitly verified field redefinition proposed in \cite{west}
in order to simplify the BRST operators for the general sequence of
$W_{2,s}$ BRST operators with $s=4$, 5, 6, and 7.  We showed that after the
non-local field transformation (\ref{dis}) and some further field
redefinitions, all these BRST operators can be transformed into an
$s$-independent form (\ref{tr}), whose cohomology is trivial.  For the
exceptional $W_{2,s}$ BRST operators, it is unclear whether the above
procedure can be carried out.

    It was recently proposed that the bosonic string could be viewed as a
special vacuum of the $N=1$ superstring, which could in turn be viewed as
a special vacuum of the $N=4$ superstring \cite{vafa}.  It was also suggested
that the bosonic string might be viewed as the lowest member of a hierachy
of $W_N$ strings \cite{vafa,figo}.  In fact, as observed in \cite{zhao}, there
is a sense in which this is already seen in the usual multi-scalar $W_3$
string,
where physical states are those of $c=25\ft12$ bosonic strings tensored with
the Ising model.  Another proposal was made recently \cite{west}, in which
a sequence of field transformations was made in order to arrive at a theory
with the cohomology of the ordinary bosonic string, starting from the $W_3$
string.  However, one step along the sequence involved the non-local
transformation (\ref{dis}), which yields the form (\ref{pf}) for $Q_1$ and, as
we have seen, gives an empty cohomology for $Q_B$. Although
in the subsequent transformations in \cite{west} a theory with the cohomology
of the bosonic string is obtained, it would seem that one should not interpret
the result as showing that this bosonic string is a special vacuum of the
original $W_3$ string, since at
an intermediate stage all the cohomology was lost. It would be interesting
to see whether there is any way to bypass the step (\ref{dis}), and find a
non-singular projection from the states of the $W$ string to
those of the bosonic string.

\vspace{1.5truecm}
\noindent{\it {\Large Acknowledgements}}
\vspace{1.0truecm}

We should like to thank Horst Kausch for helpful discussions.  We have made
extensive use of the Mathematica package OPEdefs \cite{ope} written by K.
Thielemans.

\end{document}